\documentclass[twocolumn]{aastex701} %

\usepackage{amsmath} 
\usepackage{tabularx,booktabs} 
\usepackage{gensymb} 
\usepackage{subcaption} 

\usepackage{xcolor}
\usepackage{float}  
\usepackage{braket}

\usepackage[normalem]{ulem}
\usepackage[dvipsnames]{xcolor}

\begin{document}

\title{
Sometimes You Just Can't Put a Ring on It: \\
Setting Constraints on Rings around Moons from Magnetic Fields
}

\author[orcid=0009-0007-5992-4151,sname='Erak']{Jamie M. Erak}
\affiliation{Department of Physics and Astronomy, Curtin University, Perth, WA 6102, Australia}
\affiliation{Institute of Astronomy, University of Cambridge, Madingley Road, Cambridge CB3 0HA, UK}
\email[show]{jamie.erak@curtin.edu.au}

\author[orcid=0000-0002-2728-0132,sname='Rozner']{Mor Rozner} 
\affiliation{Institute for Advanced Study, Einstein Drive, Princeton, NJ 08540, USA}
\affiliation{Institute of Astronomy, University of Cambridge, Madingley Road, Cambridge CB3 0HA, UK}
\affiliation{Gonville \& Caius College, Trinity Street, Cambridge CB2 1TA, UK}
\email{morozner@ast.cam.ac.uk}

\begin{abstract}
All four giant planets and several minor bodies in the Solar System host rings. However, rings around moons have yet to be observed.
A host planet can produce magnetic fields that affect its moons, adding a wealth of dynamical phenomena that could shape the properties of such ring systems.
In this study, we investigate constraints on the stability of circumsatellital rings (CSRs) under the effect of magnetic fields originating from the
host
planet, using both analytical and numerical methods. 
We find that the electric field induced by the rotation of the ambient planetary magnetosphere constitutes a significant perturbation on charged grains in CSRs. We demonstrate that this effect can de-orbit sufficiently charged grains on short timescales, providing a novel approach to constrain the properties of CSRs.
\end{abstract}

\keywords{\uat{Planetary rings}{1254} --- \uat{Natural satellites (Solar system)}{1089} --- \uat{Magnetic fields}{994}}


\section{Introduction} 

Ring systems are some of the most fascinating and complex  structures found in our Solar System, and constitute a fertile ground for various dynamical phenomena
\citep[see detailed reviews in][]
{Goldreich_1982,Esposito2006,
Tiscareno_2013}. 
The study of planetary ring systems is also a valuable tool for understanding the dynamics of other disk systems, such as accretion disks and spiral galaxies, as they can be probed by spacecraft \citep{Tiscareno_2013}. Processes that can lead to the formation of ring systems, which include grazing collisions and close encounters with comets and asteroids \citep{Charnoz_2018}, commonly occur throughout the Solar System. In addition to the gas giants, several minor bodies in the Solar System have been found to possess rings, including the Centaurs Chariklo \citep{Braga-Ribas_2014} and Chiron \citep{Ortiz_2015} and the trans-Neptunian dwarf planets Haumea \citep{Ortiz_2017} and Quaoara \citep{Morgado_2023}. 
The Solar System hosts more than $400$ moons\footnote{See \url{https://ssd.jpl.nasa.gov} for an updated count.} with various properties, most of them orbiting giant planets.
Despite this, there is no current evidence that any moons in the Solar System possess rings, which we will refer to as circumsatellital rings (CSRs). 

Whether such CSRs once existed and have later decayed remains uncertain, but this scenario could explain certain orbital and surface features of some moons in the Solar System. One interesting example is the huge equatorial ridge of Iapetus, which could have formed due to a collapsed ring system \citep{Ip_2006}.
Understanding the parameter space available for
CSRs in the Solar System is also highly relevant for extrasolar systems. While exomoons are yet to be detected, ringed exomoons (also referred as `cronomoons') would be easier to detect using the transit method due to their larger apparent size, and \cite{Sucerquia_2022} found that, for moons of exoplanets in close proximity to their host star, such CSRs could both form and survive for long enough to be detected.
Recently, 
\cite{Sucerquia_2024},
studied the stability of hypothetical ring systems around Solar System moons using N-body simulations to investigate the perturbing influence of the host planet and companion satellites on the orbits of test particles around $18$ moons. It was concluded that dynamical stability considerations alone cannot rule out the existence of CSRs. Hence, their current absence in the Solar System remains an open question, likely attributed to non-gravitational phenomena, such as stellar radiation,  magnetospheric drag, and/or magnetic fields.

The latter is of particular interest, since moons in the outer Solar System are immersed in complex magnetic environments, and there are a variety of processes by which grains in a ring acquire charge \citep{Graps_2008}. It has been shown \citep{Burns1980,Morfill1980,
Northrop_1982} that electromagnetic fields play an important role in shaping the structure of planetary rings. Several moons in the Solar System have induced magnetic fields due to subsurface conducting layers, such as oceans \citep{Kivelson_2000}, and Ganymede possesses both an induced and an internal magnetic field \citep{Kivelson_2002}. In the context of CSRs, however, the dominant magnetic field originates from the host planet, as the gas giants of the outer Solar System all possess strong and rapidly rotating magnetic fields.

In this study, we focus on the influence of the host planet's magnetosphere on the dynamics and stability of charged grains in CSRs, and provide a novel approach to set constraints on the available parameter space of CSRs. We begin by numerically solving the equation of motion of a charged grain in a CSR. We then introduce an analytical model to describe the evolution of grain orbits, and derive a formula constraining the orbits of grains for a given charge-to-mass ratio. We conclude with a discussion of the implications for CSR stability.

\section{Numerical Model} \label{Num}

Understanding the orbits of charged particles subject to both electromagnetic and gravitational forces is a difficult mathematical task. Analytic results can be obtained only for cases involving a high degree of symmetry: for instance, using Hamiltonian theory, \cite{Northrop_1982} derived a marginal stability radius for negatively charged grains orbiting in Saturn's ring plane, within which a perturbed grain will leave the ring plane and strike Saturn's surface. CSRs are a more complex problem, however, and in general, one will have to resort to approximations and/or numerical integration.

In SI units, the Newtonian equation of motion of a particle subject to both electromagnetic and gravitational forces is
\begin{equation} \label{EoM}
    \frac{d\vec{v}}{dt}=\frac{q}{m}\left(\vec{v}\times\vec{B}+\vec{E}\right)+\vec{g},
\end{equation}
where $q/m$ is the charge-to-mass ratio, $\vec{B}$ is the magnetic field, $\vec{E}$ is the electric field, and $\vec{g}$ is the gravitational field. Here, we apply Eq.~(\ref{EoM}) to a charged grain in a CSR. We assume that the host planet and the moon are spherically symmetric, and neglect gravitational perturbations from other satellites or any other objects in the vicinity.

\begin{figure}[ht!]
\centering
\includegraphics[width=\linewidth]{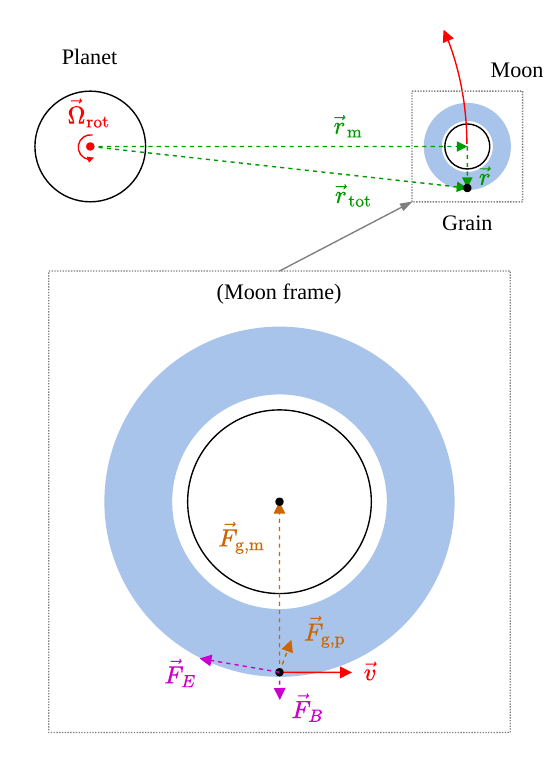}
\caption{Schematic diagram (not to scale) of a planet, a moon, and a grain in a CSR (shaded region). The inset illustrates the forces acting on a grain in the moon's frame: the electric force ($\vec{F}_E$), the magnetic force ($\vec{F}_B$), the gravitational force from the moon ($\vec{F}_{\rm g,m}$), and the gravitational perturbation due to the host planet ($\vec{F}_{\rm g,p}$).
\label{fig:CSR}}
\end{figure}
We also ignore the effects of any internal satellitial magnetic fields in the moons, so that $\vec{B}$ is the magnetic field of the host planet (assumed to be an axisymmetric dipole field). The geometry of the problem is illustrated in Fig.~\ref{fig:CSR}.

In a rotating magnetosphere, the electric field is given by \citet{Birmingham_1979} as 
\begin{equation} \label{E}
    \vec{E}=\vec{B}\times\vec{v}_{\rm plasma},
\end{equation}
where $\vec{v}_{\rm plasma}$ is the flow velocity of the magnetospheric plasma. A common approximation is to assume perfect corotation of the plasma with the planet, i.e., $\vec{v}_{\rm plasma}=\vec{\Omega}_{\rm rot}\times\vec{r}_{\rm tot}$, where $\vec{\Omega}_{\rm rot}$ is the angular velocity of the planet's rotation and $\vec{r}_{\rm tot}$ is the grain's position vector relative to the centre of mass of the host planet. While this is valid at distances very close to the host planet, the magnetospheric plasma deviates from full corotation at larger distances due to radial transport and dissipative effects. 
To describe the induced electric field (\ref{E}) in our models, we assume that the azimuthal component of the plasma velocity is dominant, but that it may rotate at sub-corotation angular velocities:
\begin{equation} \label{E2}
    \vec{E}=\vec{B}\times(\vec{\Omega}_{\rm plasma}\times\vec{r}_{\rm tot}).
\end{equation}

The value of $\Omega_{\rm plasma}$ varies from one system to another. The effect discussed here is general, but we will focus on parameters taken from our Solar System.
Near Rhea, the plasma velocity is about $70\%$ of what would be expected from full corotation \citep{Wilson_2010}, so we set $\Omega_{\rm plasma}=0.7\,\Omega_{\rm rot}$. The velocity profile of \cite{Saur_2004} suggests that the azimuthal plasma velocity flattens out to about $120\,\rm{km/s}$ above $20$ Saturn radii. At the distance of Titan, this is about $60\%$ of what would be expected from full corotation, giving $\Omega_{\rm plasma}=0.6\,\Omega_{\rm rot}$. During the times that Iapetus is within Saturn's magnetosphere, this is about $20\%$ of full corotation, giving $\Omega_{\rm plasma}=0.2\,\Omega_{\rm rot}$.

To numerically solve the equation of motion (\ref{EoM}) for a single charged grain, we made use of the \texttt{REBOUND} N-body code \citep{Rein_2012} with an additional velocity-dependent force defined to incorporate the Lorentz force. Grains were initialised in circular orbits ($e=0$) around the host moon, assumed to be coplanar and corotating with the moon's orbit. Here, the integration duration was chosen to be just one orbital period of the moon around the planet ($T_{\rm m}$), since the electromagnetic perturbation acts on short timescales and is periodic.
It should be noted that while we focus mainly on short-term effects in the present study, which is sufficient to determine whether the electromagnetic perturbation can be dynamically significant, effects could accumulate. We discuss secular evolution in more detail in \ref{subsub: secular evolution}.

\begin{figure}[ht!]
\centering
\includegraphics[width=\linewidth]{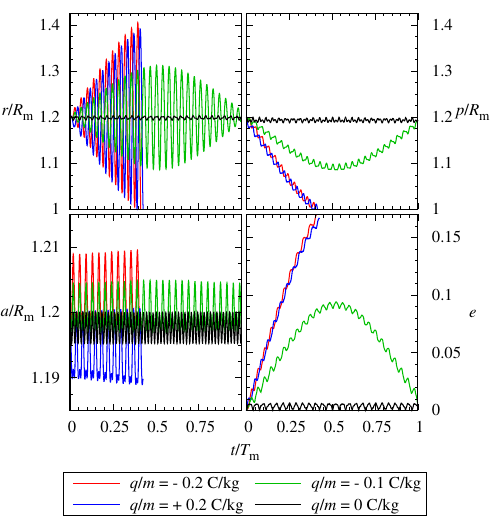}
\caption{Separation distance from moon ($r$), periapsis distance ($p$), semimajor axis ($a$), and eccentricity ($e$) plotted as functions of time for a grain orbiting Rhea. Our numerical results are presented for grains with charge-to-mass ratios of $-0.2$, $-0.1$, $0$, and $+0.2\,\rm{C/kg}$. \label{fig:Rhea_ts}}
\end{figure}

Fig.~\ref{fig:Rhea_ts} presents simulation results for grains in orbit of Rhea, showing the evolution of
the distance of the grain from the moon's centre of mass ($r$), the periapsis distance ($p$), semimajor axis ($a)$, and eccentricity ($e$) over one orbital period of Rhea around Saturn. The initial orbital radius was chosen to be $1.2\,R_{\rm m}$, where $R_{\rm m}$ is the radius of the host moon, and results are presented for four different choices of charge-to-mass ratios: $0, \ \pm 0.2, \ -0.1 \ \rm{C/kg}$. These values were chosen to showcase the various dynamical regimes (stable orbits or collisions). It should be noted that the value of $q/m$ varies widely, and one of the aims of this study is to explore the available parameter space and how different choices affect stability; see further discussion in Section~\ref{properties}.

For $q/m=0\,\rm{C/kg}$, electromagnetic forces are not present, allowing us to isolate the effect of the gravitational perturbation from Saturn. For $-0.1$ and $\pm0.2\,\rm{C/kg}$, the electromagnetic perturbation becomes dominant, and the orbit deviates significantly from circular. The distance (top left panel) shows oscillations with a period close to that of the grain's orbit around Rhea, in addition to a slower modulation with a period close to that of Rhea's orbit around Saturn. This slow modulation is consistent with the evolution of the eccentricity. The electromagnetic perturbation also excites oscillations in the semimajor axis, although these are small in magnitude and do not cumulate. The changing eccentricity is the primary driver of the evolution of the periapsis distance.

For $q/m=-0.2\,\rm{C/kg}$, the electromagnetic perturbation is large enough that the separation distance falls below the radius of Rhea, indicating that the grain strikes the surface within a short timescale. This occurs at approximately $0.4\,T_{\rm m}$, or $43$ hours, which is about the same time that the periapsis distance falls below $R_{\rm m}$. We also include a positively charged grain with charge-to-mass ratio $0.2\,\rm{C/kg}$ to demonstrate that the effect on the orbital parameters is almost identical. The perturbation from the electric field has opposite phase (and indeed, we can observe that the separation distance oscillates with opposite phase), which leads to small differences in the eccentricity and periapsis distance due to interference with the gravitational perturbation.

Simulation results for grains in orbit of Titan and Iapetus are included in Appendix \ref{A1}. Qualitatively, the behaviour is very similar to what we observed for Rhea, but the effect of the gravitational perturbation from Saturn becomes negligible. The oscillations in the separation distance also become harder to resolve, since the period of the moon's orbit around Saturn becomes much greater than the period of the grain's orbit around the moon. The change in the semimajor axis becomes insignificant (i.e., no net work is done on the grain), and the difference in the dynamics of negatively and positively charged grains becomes vanishingly small.

\section{Analytical Model} \label{ana}

In this section, we derive the constraints from 
planetary magnetic fields using an analytical approach. We start with the Gauss planetary equation for the eccentricity of the grain's orbit around the host moon \citep[e.g.,][]{MurrayDermott1999}:
\begin{equation} \label{Gauss}
    \frac{de}{dt}=\sqrt{\frac{a(1-e^2)}{GM_{\rm m}}}\left[F_{r}\sin{\theta}+F_\theta(\cos{\theta}+\cos{\mathcal{E}})\right],
\end{equation}
where $\theta$ is the true anomaly, $\mathcal{E}$ is the eccentric anomaly, and $F_r$ and $F_\theta$ are respectively the radial and angular components of the perturbing force per unit mass. Gravity-associated stability analysis was performed in previous studies \citep[e.g.,][]{Sucerquia_2024}, and we focus purely on the effects arising from magnetic fields. Furthermore, for the moons presently studied, it is valid to assume that the magnetospheric plasma velocity, $\vec{v}_{\rm plasma}=\vec{\Omega}_{\rm plasma}\times\vec{r}_{\rm tot}$, is much greater than the grain's velocity in the moon's frame. As such, we consider only the perturbing force from the electric field in the host planet's magnetosphere, given by Eq.~(\ref{E2}). In the moon's frame, this becomes
\begin{equation} \label{E3}
    \vec{E}=\vec{B}\times
    \left[(\vec{\Omega}_{\rm plasma}-\vec{\Omega}_{\rm m})\times\vec{r}_{\rm tot}\right],
\end{equation}
where $\vec{\Omega}_{\rm m}$ is the angular velocity of the moon's orbit around the host planet. We assume that the moon's orbit is circular and equatorial, that the grain's orbit is coplanar, and that the magnetic field is a dipole field aligned with the host planet's rotation axis. We also assume that the distance from the grain to the moon is much smaller than the distance $r_{\rm m}$ from the moon to the planet, so that $\vec{r}_{\rm tot}\approx \vec{r}_{\rm m}$. Eq.~(\ref{E3}) then gives
\begin{equation}
    |\vec{E}|=\frac{\mu_{\rm p}}{r_{\rm m}^2}(\Omega_{\rm plasma}-\Omega_{\rm m}),
\end{equation}
which is directed along $\hat{r}_{\rm m}$. Here, $\mu_{\rm p}$ is the magnetic dipole moment of the host planet, and $\Omega_{\rm m}=\sqrt{GM_{\rm p}/r_{\rm m}^3}$, where $M_{\rm p}$ is the mass of the host planet. The perturbing force per unit mass therefore satisfies
\begin{equation} \label{magnitude}
    |\vec{F}|=\left|\frac{q}{m}\right|\frac{\mu_{\rm p}}{r_{\rm m}^2}(\Omega_{\rm plasma}-\Omega_{\rm m}),
\end{equation}
which for negatively charged grains is directed inward from the grain to the host planet. We describe the position of the grain relative to the moon with a Cartesian coordinate
system, with the $\hat{y}$ axis aligned with the periapsis. At $t=0$, we assume the $\hat{x}$ axis is aligned with $\hat{r}_{\rm m}$. The electric force vector shifts around the moon with angular velocity $\Omega_{\rm m}$, and to enforce the periodicity of $2T_{\rm m}$ observed for the eccentricity in Section~\ref{Num}, we assume that the periapsis shifts around the moon with angular velocity $\Omega_{\rm m}/2$. As such, the direction of the perturbing force on the grain is
\begin{equation}
    \hat{F}=-\cos\left(\frac{\Omega_{\rm m}}{2}t\right)\hat{x}-\sin\left(\frac{\Omega_{\rm m}}{2}t\right)\hat{y}.
\end{equation}
Transforming to a polar coordinate system with $\theta$ measured from the $\hat{y}$ axis, we have
\begin{equation}
    \hat{F}=\sin\left(\theta-\frac{\Omega_{\rm m}}{2}t\right)\hat{r}+\cos\left(\theta-\frac{\Omega_{\rm m}}{2}t\right)\hat{\theta}.
\end{equation}
Motivated by our numerical results in Section~\ref{Num}, we consider the semimajor axis $a$ as constant. We now expand Eq.~(\ref{Gauss}) to zeroth order in eccentricity. The Keplerian relations then yield $\theta=M+2e\sin M+\mathcal{O}(e^2)$ and $\mathcal{E}=M+e\sin M+\mathcal{O}(e^2)$ \citep{MurrayDermott1999}, where $M=\Omega_{\rm g}t$ is the mean anomaly and $\Omega_{\rm g}=\sqrt{GM_{\rm m}/a^3}$ is the mean motion of the grain around the moon. Substituting the zeroth order of these expansions, i.e. $\theta\approx \mathcal{E} \approx M$, into Eq.~(\ref{Gauss}), it reduces to 
\begin{align}\label{eq:dedt}
    \frac{de}{dt}=\frac{|\vec{F}|}{2}\sqrt{\frac{a}{GM_{\rm m}}}\bigg[&3\cos\left(\frac{\Omega_{\rm m}}{2}t\right) \nonumber \\
    &+\cos\left(\bigl\{2\Omega_{\rm g}-\Omega_{m\rm}/2\bigr\} t\right)\bigg],
\end{align}
which can be integrated analytically, giving
\begin{align} \label{sol}
    e(t)=\frac{|\vec{F}|}{2}\sqrt{\frac{a}{GM_{\rm m}}}\bigg[&\frac{6}{\Omega_{\rm m}}\sin\left(\frac{\Omega_{\rm m}}{2}t\right) \nonumber \\
    &+\frac{\sin\left(\big\{2\Omega_{\rm g}-\Omega_{\rm m}/2\big\}t\right)}{2\Omega_{\rm g}-\Omega_{\rm m}/2}\bigg].
\end{align} 
Assuming that $\Omega_{\rm g}\gg\Omega_{\rm m}$, the second term vanishes, leaving
\begin{equation} \label{sol2}
    e(t)=\frac{3|\vec{F}|}{\Omega_{\rm m}}\sqrt{\frac{a}{GM_{\rm m}}}\sin\left(\frac{\Omega_{\rm m}}{2}t\right),
\end{equation}
which has a maximum at $t=T_{\rm m}/2$. We note that this solution reproduces the expected periodicity, and that changing the periapsis shift of $\Omega_{\rm m}/2$ postulated earlier would change the frequency and amplitude, but not the form of the solution.

Our goal is to determine the critical initial orbital radius $r_{\rm c}$ below which a grain of given $|q/m|$ will collide with the moon, analogous to the marginal stability radius considered by \cite{Northrop_1982}. The periapsis distance,
\begin{equation}
    p(t)=a[1-e(t)],
\end{equation}
has a minimum when $e(t)$ has a maximum, and setting this minimum value equal to the moon radius $R_{\rm m}$ yields a relation between $r_{\rm c}=a$ and $|q/m|$. After some algebra, we obtain
\begin{equation} \label{analytic}
    \left|\frac{q}{m}\right|=\frac{G}{3\mu_{\rm p}(\Omega_{\rm plasma}-\Omega_{\rm m})}\sqrt{\frac{M_{\rm p}M_{\rm m}r_{\rm m}}{r_{\rm c}}}\left(1-\frac{R_{\rm m}}{r_{\rm c}}\right).
\end{equation}

\begin{figure}[hb!]
\centering
\includegraphics[width=\linewidth]{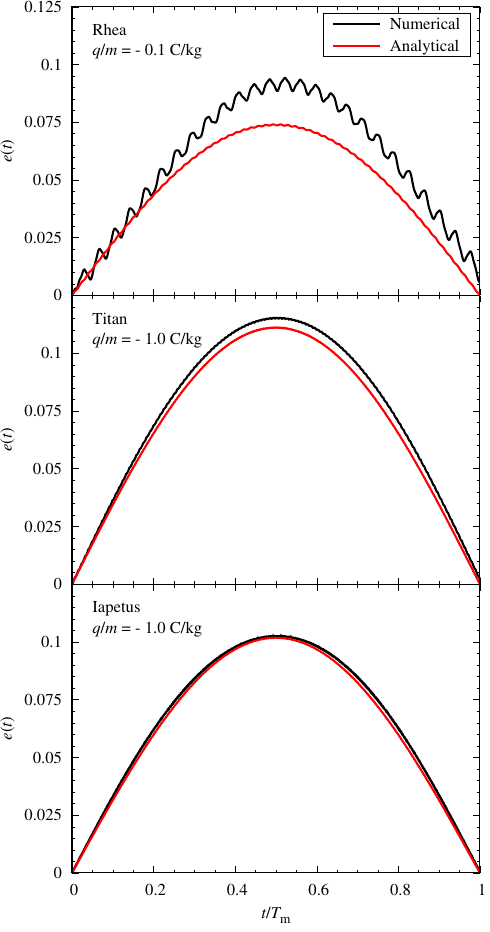}
\caption{The eccentricity of a grain with initial orbital radius $1.2\,R_{\rm m}$ plotted as a function of time for Rhea (top), Titan (mid), and Iapetus (bottom). For Rhea, the charge-to-mass ratio was chosen to be $-0.1\,\rm{C/kg}$, while for Titan and Iapetus, it was $-1.0\,\rm{C/kg}$. Our analytical results, given by Eq.~(\ref{sol}), are compared with our numerical results.
\label{fig:ts}}
\end{figure}

Eq.~(\ref{analytic}) can be thought of as the critical value of $|q/m|$ above which a grain of given initial orbital radius $r_{\rm c}$ will collide with the moon. We can plot it parametrically, however, to obtain the boundary of the region of stability in the grain's parameter space (as is done in Section~\ref{results}, where we want the critical initial orbital radius $r_{\rm c}$ as a function of $|q/m|$). As an example, for the parameters of Titan and a charge-to-mass ratio of $-1\,\rm{C/kg}$, the critical radius is approximately $2890\,\rm{km}$, which is $1.12$ times the radius of Titan and $0.700$ times the Roche radius. We can also read off some useful insights regarding the stability of CSRs: the critical $|q/m|$ is smaller (i.e., the restriction on the stability of grains is more severe) for host planets with strong magnetic dipole moments and rapidly rotating magnetospheres, and for small moons orbiting close to the host planet.

The derivation presented here made several approximations: dominance of the electric perturbation, circular and equatorial moon orbits, coplanar grain orbits, aligned dipole fields, $\vec{r}_{\rm tot}\approx\vec{r}_{\rm m}$, low-order eccentricity expansion, constant semimajor axis, and the assumed periapsis evolution to reproduce the expected periodicity. In what follows, we investigate its validity for the three moons presently considered by comparison with numerical results.

In Fig.~\ref{fig:ts}, we plot the evolution of the eccentricity described by Eq.~(\ref{sol}) for charged grains in orbit of Rhea, Titan, and Iapetus. The initial orbital radius was chosen to be $1.2\,R_{\rm m}$, and the charge-to-mass ratio was chosen to be $-0.1\,\rm{C/kg}$ for Rhea and $-1.0\,\rm{C/kg}$ for Titan and Iapetus. For each moon, we find reasonably good agreement with our numerical results, demonstrating the validity of the analytical model.

For Rhea, the numerical solution exhibits fast oscillations due to the gravitational perturbation from Saturn, which is not accounted for in the analytical solution. This causes the numerical solution to deviate from the analytical solution, with the eccentricity becoming larger. This effect is reduced for Titan, and is negligible for Iapetus,
as indeed expected due to a weaker gravitational effect at larger orbital distances. 

\begin{figure}[ht!]
\centering
\includegraphics[width=\linewidth]{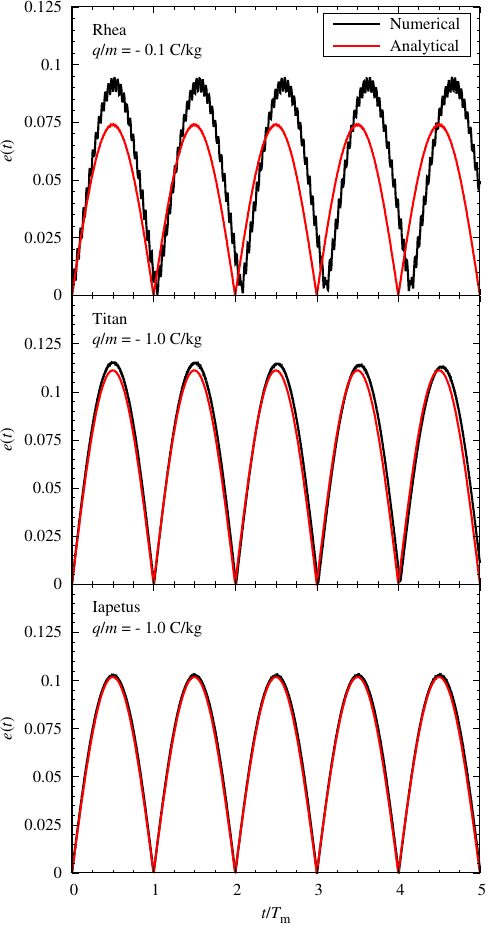}
\caption{Same as Fig.~\ref{fig:ts}, but with the solutions presented up to $t=5T_{\rm m}$. To extend it to $5T_{\rm m}$, we took the absolute value of the analytical solution.
\label{fig:ts2}}
\end{figure}

Fig.~\ref{fig:ts2} shows the evolution of the grain's eccentricity up to $t=5T_{\rm m}$, comparing our numerical and analytical results (the analytical solution was extended to $5T_{\rm m}$ by taking the absolute value). We see that the long-term behaviour is almost identical, with the perturbation consisting of periodic oscillations in $e(t)$. We again note excellent agreement between the numerical and analytical solutions for Iapetus. This deteriorates slightly for Titan, and more noticeably for Rhea, since they orbit nearer to Saturn. In the numerical solutions for the latter two, it can be observed that the period is modified. We expect this drift is due to coupling with the gravitational perturbation from Saturn, which has periodicity comparable to that of the grain's orbit around the moon. For moons orbiting near their host planet, this becomes significant compared to the period of the moon's orbit around the planet, affecting the period of the total perturbation.

\section{Constraints on CSRs} \label{results}

We now turn our attention to the constraints placed on the parameter space of grains in CSRs within our numerical and analytical models. To estimate an upper bound for the initial orbital radius of a grain in a CSR, we used the Roche limit, which is defined as the separation distance at which a body’s gravitational self-attraction is exceeded by the tidal forces from a primary body (in our case, the host moon):
\begin{equation}
    r_{\rm R}=R_{\rm m}\left(\frac{2\rho_{\rm m}}{\rho_{\rm g}}\right)^{1/3}
\end{equation}
where $\rho_{\rm m}$ is the density of the moon and $\rho_{\rm g}$ is the density of the CSR material. Following \cite{Sucerquia_2024}, we estimate $\rho_{\rm g}$ to be that of ice formed from water ($917\,\rm{kg/m}^3$). Although it is not an exact boundary, the Roche limit can be considered as a guide
for the outer radius of a ring system, as material orbiting beyond it will tend to coalesce into larger objects. The inner radius of the ring system is taken to be the moon radius, $R_{\rm m}$.

Another relevant lengthscale is the Hill radius, which describes the extent of the region where the gravitational pull of a secondary body (the host moon) dominates over that of a primary body (the host planet). It is given by
\begin{equation}
    r_{\rm H}=a_{\rm m}(1-e_{\rm m})\sqrt[3]{\frac{M_{\rm m}}{3M_{\rm p}}},
\end{equation}
where $a_{\rm m}$ and $e_{\rm m}$ respectively denote the semimajor axis and eccentricity of the moon. It was shown by \cite{Domingos_2006} that the dynamical stability limit, beyond which objects orbiting the secondary body become unstable due to the pull of the primary, is given simply by $cr_{\rm H}$, where $c\approx0.4895$ for prograde orbits. The ratio $r_{\rm R}/r_{\rm H}$ yields insights into the stability of a CSR, with a low ratio (i.e., $r_{\rm R}\ll r_{\rm H}$) signifying high stability against gravitational perturbations. We find that $r_{\rm R}/r_{\rm H}=0.182$ for Rhea, $0.0811$ for Titan, and $0.0278$ for Iapetus, all of which are well below the dynamical stability limit of $0.4895$. The particularly low value for Iapetus is consistent with the negligible gravitational perturbations from Saturn observed in Section~\ref{ana}. A more detailed analysis was carried out by \cite{Sucerquia_2024}, who found that ring particles around moons with lower Roche-to-Hill ratios are indeed more stable against gravitational perturbations, with Iapetus the most stable among the set of 18 moons considered.

The charge-to-mass ratio and initial radius of the grain’s orbit were considered free parameters, and for given values of these parameters, we assessed the stability of the grain’s orbit by determining whether it would collide with the host moon on a timescale of $1\,T_{\rm m}$. This neglects effects that occur over secular timescales, but suffices for our purpose of investigating the electromagnetic perturbation as a rapid grain-removal mechanism.
For a given charge-to-mass ratio $q/m$, we determined the critical radius $r_{\rm c}$ numerically using a bisection algorithm, and it was verified that integrations over longer timescales (up to $5\,T_{\rm m}$) led to negligible difference in $r_{\rm c}$. For the analytical results, we used Eq.~(\ref{analytic}).

\begin{figure}[ht!]
\centering
\includegraphics[width=\linewidth]{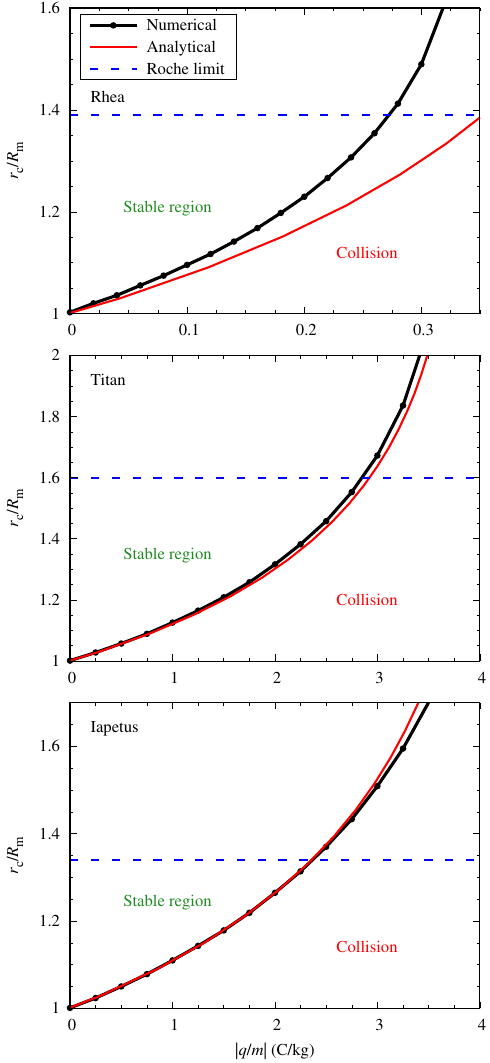}
\caption{{The critical initial orbital radius, below which a grain will crash into the moon's surface, plotted as a function of the charge-to-mass ratio for Rhea (top), Titan (mid), and Iapetus (bottom). Our analytical results, given by Eq.~(\ref{analytic}), are compared with our numerical results, and the Roche limit is indicated by a horizontal dashed line. 
}
\label{fig:stability}}
\end{figure}

In Fig.~\ref{fig:stability}, we present the critical initial orbital radius as a function of $|q/m|$ for negatively charged grains in orbit of Rhea, Titan, and Iapetus. The Roche limit ($r_{\rm R}$) is also indicated. For initial radii below $r_{\rm c}$, the grain will crash into the surface of the host moon within $1\,T_{\rm m}$, therefore the region of parameter space available to grains in the CSR is above $r_{\rm c}$ but below $r_{\rm R}$. Our numerical and analytical results are found to be in good agreement for all three moons considered.

For Rhea, the analytical $r_{\rm c}$ is uniformly smaller than the numerical result. This is consistent with the discrepancy observed in Fig.~\ref{fig:ts}, where the analytical eccentricity remains smaller than the numerical result due to the gravitational perturbation from Saturn, which is neglected in the analytical model. The qualitative behaviour is the same, however: as $|q/m|$ increases, the critical radius moves further outward until it exceeds the Roche limit. For Titan and Iapetus, remarkable quantitative agreement is found for small values of $|q/m|$. For larger $|q/m|$, where the perturbation becomes larger, a small difference arises; higher-order calculations are required in this regime to increase the accuracy of the analytical description.

One can observe that, for any given $|q/m|$, the critical radius relative to the moon radius is much smaller for Iapetus and Titan than for Rhea. This agrees with the intuition that CSRs are more stable around moons further from the host planet, where the magnetic field is weaker and the magnetospheric plasma lags further behind corotation.

The comparisons with numerical results in Figs.~\ref{fig:ts}, \ref{fig:ts2}, and \ref{fig:stability} demonstrate the range of applicability of the analytical approxination. Due to its inherent assumptions, the analytical approach introduced here is only valid for moons in equatorial and circular orbits around planets, or small perturbations from them, with magnetic fields similar to aligned dipoles. One can expect the analytical approach to be highly accurate for moons distant from the host planet, where the electric perturbation dominates over the gravitational and magnetic perturbations, as well as for grains with small charge-to-mass ratios, since the eccentricity remains small enough to be treated to low order. In other cases, the approximations described in Section~\ref{ana} lose validity, and the analytical approach should be considered only a heuristic description. 

The relative strengths of the electric and gravitational perturbation can be quantified by assessing $\xi=|\vec{F}_E|/|\vec{F}_{\rm grav}|$, where $|\vec{F}_E|$ is given by Eq.~(\ref{magnitude}) and the gravitational perturbation due to the host planet is given by 
\begin{equation}
    |\vec{F}_{\rm grav}|\approx\frac{GM_{\rm p}r}{r_{\rm m}^3},
\end{equation}
which yields
\begin{equation}
   \xi= \left|\frac{q}{m}\right|\frac{\mu_{\rm p}r_{\rm m}}{GM_{\rm p}r}(\Omega_{\rm plasma}-\Omega_{\rm m}).
\end{equation}
The electric perturbation dominates over the gravitational one when the ratio $\xi$ is bigger than one. The approximation that neglects the gravitational contribution starts to break when this ratio approaches unity. Indeed, for example, a grain with $|q/m|=0.2\,\rm{C/kg}$ and initial orbital radius $1.2\,R_{\rm m}$ will have $\xi=1.4$ for Rhea. The deviations from the approximation can be seen in Fig.~\ref{fig:stability}.

\section{Discussion}

\subsection{Ring Ages \& Decay}

Rings are not static structures, and their evolution proceeds after their initial buildup, potentially leading towards decay of the rings. Here we will briefly review the different timescales associated with these processes.

\subsubsection{Viscous Spreading}

Inner particles in the ring are expected to lose angular momentum and fall towards the hosting satellite, while the opposite holds for outer particles, leading to a spread between rings, such that the overall net angular momentum transfer is outward and the mass of the ring is transferred inward \citep[see detailed reviews in][]{Goldreich_1982,
Esposito2006}. A similar process was studied in the context of accretion discs around black holes \citep{LyndenBellPringle1974}. The viscous spreading timescale is given by   

\begin{align}
\tau_{\rm{spread}} &=\frac{(\Delta r)^2}{\nu}\approx \\ &\approx 75 \ \rm{Myr} \ \left(\frac{\Delta r}{500 \ \rm{km}}\right)^2\left(\frac{1 \ \rm{cm^2} \  \rm{sec}^{-1}}{\nu}\right),  
\end{align}

\noindent
where $\Delta r$ is the ring width and $\nu$ is the viscosity. Using $\nu_{\rm{min}}= \Omega_{\rm{Kep}}\left(\Sigma/\rho\right)^2$ where $\Omega_{\rm{Kep}}$ is the Keplerian frequency, $\Sigma$ is the surface density of the ring, and $\rho$ is the density of the ring particle. For the typical parameters discussed throughout the paper, and  $\Sigma \approx 100 \ \rm{g} \ \rm{cm}^{-2}$, the typical viscosity is of order unity, and we adopt $\nu=1 \ \rm{cm}^2 \ \rm{sec}^{-1}$. Other depletion mechanisms are expected to deplete the rings more efficiently. 

\subsubsection{Erosion}

Erosive mechanisms, such as micrometeorites and sputtering, can degrade ring particles. Impactors hit the ring and transfer kinetic energy to it, potentially fragmenting dust grains into smaller ones and changing the size distribution, similarly to the mechanisms discussed for planetary rings \citep[e.g.,][]{Northrop_1987}. 

\subsubsection{Secular Effects}\label{subsub: secular evolution}

Secular effects might accumulate over time and contribute significantly to ring depletion. In Eq.~(\ref{eq:dedt}), we derive the instantaneous eccentricity change, which we can orbit-average to get a secular equation:
\begin{align}
\Braket{\frac{de}{dt}} &=\frac{1}{2\pi }\int_0^{2\pi}\frac{de}{dt}dM \\ \nonumber 
&=\frac{|\vec{F}|}{T_{\rm m}}\sqrt{\frac{a}{GM_{\rm m}}}\frac{\sin\left(\pi\left[4\Omega_{\rm g}/\Omega_{\rm m}-1\right]\right)}{4\Omega_{\rm g}-\Omega_{\rm m}},
\end{align}
where $M=\Omega_{\rm m}t$ is here the mean anomaly of the moon's orbit around the host planet. The orbit-averaged change in eccentricity is very small for the regimes we consider, where $\Omega_{\rm g}\gg\Omega_{\rm m}$, but will accumulate over many orbits of the moon around the planet. Over a period, the grain is expected to experience an eccentricity change of order $\Delta e \approx T_{\rm m} \braket{de/dt}$, such that we expect the grain to collide with its host moon after
\begin{align} \label{secular}
N_{\rm{coll}}\approx \frac{e_{\rm{coll}}-e_0}{|\Delta e|}
\end{align}
\noindent
orbits of the moon around the planet, where $e_{\rm{coll}}=1-R_{\rm m}/a$ is the critical eccentricity for collision. A grain orbiting Iapetus with $a=1.2\,R_{\rm m}$ and $|q/m|=1\,\rm{C/kg}$ is within the stable region according to our short-term analysis, but Eq.~\ref{secular} predicts that a collision will occur after $\approx14000$ orbits, or $\approx3000$ years. The inclusion of gravitational perturbations and higher-order contributions will further affect the secular evolution.

While we focused here on coplanar orbits, secular effects in misaligned configurations can lead to various interesting effects. Misaligned rings can give rise to the von Zeipel-Lidov-Kozai (ZLK) mechanism \citep{VonZeipel1910,Lidov_1962,Kozai1962}, which induces periodic interchanges of inclination and eccentricity. Similar effects have already been discussed in the context of discs \citep[e.g.,][]{Martinetal2014} or rings around moons \citep{Sucerquia_2017}. Significant eccentricity excitations could eventually lead to enhanced depletion of the rings. It should be noted that the system we are discussing here involves additional complications; hence, a more detailed and subtle study is needed for a full understanding.

\subsection{Grain Properties} \label{properties}

In Section~\ref{results}, we saw that for each moon, there exists a value of $|q/m|$ beyond which the critical radius exceeds the Roche limit. For Rhea, this is $\approx0.28\,\rm{C/kg}$ (from the numerical results), while for Titan, it is $\approx2.8\,\rm{C/kg}$, and for Iapetus, it is $\approx2.4\,\rm{C/kg}$. If the Roche limit is treated as the outer limit of the ring system, then any grain that becomes charged beyond this value will be lost from the ring, regardless of the initial position. These charge-to-mass ratios are reasonably small, and we can use the field emission limit to estimate the sizes of the grains lost from the ring.

From \cite{Mendis_1974}, the field-emission-limited potential of a sphere (in Volts) is approximately $|\phi|=910a$,
where $a$ is the radius of the sphere in $\mu$m. This leads to a restriction on $|q/m|$, which is given by \cite{Northrop_1982} as
\begin{equation} \label{limit1}
    \left|\frac{q}{m}\right|\le \frac{24.15}{ad},
\end{equation}
where $d$ is the specific gravity in $\rm{g}/\rm{cm}^{3}$. For a grain of a given charge-to-mass ratio, assumed to be spherical, Eq.~(\ref{limit1}) can be rearranged to constrain the radius according to
\begin{equation} \label{limit2}
    a\le \frac{24.15}{d}\left|\frac{q}{m}\right|^{-1},
\end{equation}
where the equality holds if the grain is charged to the field emission limit. For $d=1\,\rm{g/cm}^3$ and $|q/m|=2.8\,\rm{C/kg}$, we find that the upper limit on $a$ is about $8.6\,\mu\rm{m}$. Grains of these sizes may seem small, but are known to exist in abundance in ring systems \citep[e.g.,][]{Showalter_1987}. Indeed, this size limit is far less restricting than for grains electromagnetically removed from Saturn's rings in the erosion model of \cite{Northrop_1987}, which must be broken down to submicron sizes via micrometeorite bombardment to acquire the necessary charge-to-mass ratios (on the order of $10^2\,\rm{C/kg}$). Assuming that the composition of CSRs is similar, it follows that a broader population of grains are susceptible to electromagnetic removal after acquiring charge than for planetary rings. 

Here, we have treated the charge-to-mass ratio as a free parameter, to probe the available parameter space. Its range of values is related to the grain size via the field emission limit, and both are dynamic and interdependent. The charge-to-mass ratio of a grain depends on the charging environment and may change with time. Additionally, since smaller grains are preferentially removed, the distribution of grain sizes in a CSR can change with time. This may affect the expected signal in reflected light: for instance, the size distribution affects the scattering behaviour in polarimetry models \cite[e.g.,][]{Veenstra_2025}, which could be relevant for future applications to ringed exomoons.

\subsection{Differences from Planetary Rings}
The evolution of CSRs differs fundamentally from that of planetary rings; not only is the gravity that CSR grains experience different, but there is an essential difference in the geometry. Since the magnetic field originates from a third body rather than the hosting object of the ring, the electromagnetic perturbation on CSRs is inherently asymmetric along the orbit, compared to planetary rings, where the magnetic field originates at the hosting body. This qualitative difference leads to the different functional forms of Eq.~(\ref{analytic}) and the analogous relation derived by \cite{Northrop_1982} for planetary rings,
\begin{equation}
    r_{\rm c}^3=\frac{2}{3}\frac{GM_{\rm p}}{\left(\Omega_{\rm rot}-\frac{GM_{\rm p}}{3\mu_{\rm p}}\frac{m}{q}\right)^2}.
\end{equation}
For CSRs, we found that removal occurs for $|q/m|\sim1\,\rm{C/kg}$ typically, whereas \cite{Northrop_1982} found that for Saturn, it only becomes significant for $10^2$--$10^3\,\rm{C/kg}$. Moreover, $r_{\rm c}$ is bounded as $q/m\to-\infty$  for planetary rings, but not for CSRs, where the critical radius will grow to exceed the size of the ring system as seen in Fig.~\ref{fig:stability}. As such, we conclude that electromagnetic removal due to the ambient planetary magnetosphere is a highly significant mechanism for CSRs. It should be noted that if the moon has its own magnetic field, the process will be more similar qualitatively to the one in planetary rings. 

\subsection{Distance from the host star}
The distance from the star affects grain removal in a non-trivial manner, and there are two competing effects that the balance between them determines the result. On the one hand, closer to the star, irradiation is stronger, since the stellar UV flux scales as $d^{-2}$, where $d$ is the distance from the star. This could potentially drive the grains towards higher $q/m$ ratios (as can be seen, for example, in Fig. \ref{fig:stability}), and hence higher critical radii and larger parameter space available for CSRs. On the other hand, closer to the star, the magnetosphere could be compressed, leading to weaker perturbations and more difficult removal. We leave a more detailed discussion on these effects for future studies.
\subsection{Satellite-Sourced Particles}
In addition to other ring formation mechanisms such as grazing collisions, some moons may actively supply particles to their surroundings through, for example, volcanic activity. The broader dynamics of satellite-sourced particles in planetary magnetospheres is of high relevance in Solar System science and the search for exomoons. It has been shown that submicron-sized dust grains originating in Io's volcanic plumes can become rapidly charged due to the ambient plasma, in which case their dynamics is dominated by the Lorentz force \citep{Johnson_1980}, and these dust grains have been identified as a source of the Jovian dust streams \citep{Graps_2000}. Furthermore, dust particles created by cryovolcanism on Enceladeus are thought to be the dominant source of Saturn's E ring \citep{Spahn_2006}. More recently, the observation of a sodium transient at the hot Saturn WASP-49 A b \citep{Oza_2024} could indicate the presence of a volcanic natural satellite. These measurements provide a compelling exomoon candidate, although the origin of the transient sodium signature is yet to be pinpointed.

\subsection{Caveats and Future Directions} \label{Caveats}

\begin{itemize}
    \item We focused on magnetic fields originating from the host planet, but a similar analysis, with the proper modifications, could be carried out if the magnetic field originates in the moon itself.  
    \item We consider a single grain in our analysis here, following the approach used by \cite{Northrop_1982}, and extending it to magnetic fields induced by the host planet onto the moon. The electric and magnetic fields act on each grain separately, but further collective effects might change the picture, and are left for future studies. 
    \item The grain charging mechanisms are not well understood, even in the simpler case of planetary rings. Multiple processes are relevant, including electron and ion capture, photoelectron emission, and secondary electron emission, and these currents interact in complex ways \citep{Graps_2008}. In this study, we assumed that the charge-to-mass ratios of grains are constant, but variable charge is a more realistic scenario, and for the case of planetary rings, \cite{Jontof-Hutter_2012b} found that this significantly reduces grain stability.
    \item While we included dependence on grain composition indirectly, we expect further compositional dependencies, for example, different materials are expected to have different charging mechanisms, due to different levels of conductivity and internal structure, such as porosity and surface structure.
    \item In realistic systems, grain charging might also be affected by stellar irradiation, photoelectric effects and stellar wind, which were neglected in the present study. These effects introduce a dependence of grain charging, and therefore the effectiveness of the proposed decay mechanism, on the planet-star distance, which may impose additional constraints on the stability of ring systems.
    \item The magnetic field of the host planet was treated as an aligned dipole field. This is a good approximation in several cases \citep{Jontof-Hutter_2012b}, and especially for Saturn, where the dipole tilt is less than $0.01\degree$. A more general model would be beneficial for application to other systems (e.g., Jovian moons), however, since non-axisymmetric terms could further reduce the stability of charged grains. We also assumed a greatly simplified model of the magnetospheric plasma flow, and did not consider radial transport or the effect of the solar wind.
    \item In this study, grains were initialised in coplanar orbits, assuming circular and equatorial moon orbits. The Lorentz force depends strongly on geometry, however, and for inclined grain orbits, the perturbation may be redistributed between the radial, tangential, and vertical directions. As such, inclined rings could experience different qualitative behaviour and decay timescales.
    \item Here, we focused on the interaction between a ring particle, a moon, and the host planet, but nearby satellites can further perturb the system \citep[see][for further discussion]{Charnoz_2018,Sucerquia_2024}.
    \item While we considered in our analysis several orbits of evolution (up to $t=5T_{\rm m}$), secular effects might operate on longer timescales and produce accumulated effects that are currently not included in our approach. The present results establish that the electromagnetic perturbation can act as a rapid grain-removal mechanism, but a full secular study of the problem is an important direction for future work to establish how this effect maps into the long-term evolution of ring systems.
    
\end{itemize}

\section{Summary}

In this paper, we have investigated the dynamics of charged particles in hypothetical circumsatellital rings (CSRs) subject to magnetic fields, in order to place constraints on their stability. To treat grain dynamics in the ambient planetary magnetosphere, we utilised two approaches: numerical solution of the Newtonian equation of motion, and an analytical solution of the Gauss planetary equations (subject to some simplifying assumptions). The latter approach yielded a simple, easily applicable formula for the boundary of the stability region in the grain parameter space, providing useful physical insights into the dependence on various parameters. Both approaches demonstrate that sufficiently charged grains are lost from CSRs on short timescales due to the electric field in the host planet's magnetosphere, showing that the electric perturbation is highly significant. Moreover, we argue that electromagnetic removal events are relatively more common in CSRs than in planetary rings, since the charge-to-mass ratios at which removal occurs are smaller than for planetary rings.

This supports a scenario in which CSRs decay via a mechanism similar to the erosion model introduced by \cite{Northrop_1987} for Saturn's rings. In this model, mass is broken down into submicron-sized grains by micrometeorite impacts. The grains then become charged by plasma clouds produced by subsequent micrometeorite impacts, and are lost from the ring if within the critical radius. Our results also suggest that CSRs are more stable, and therefore more likely to be detected in extrasolar systems, if the host planet has a weak and slowly-rotating magnetosphere, the moon is large and situated farther from the host planet, or the rings consist primarily of larger objects. An inverse interpretation of this result is that the detection of rings around a moon could point to these characteristics, therefore providing insights into the planetary magnetosphere and satellite system. 

We emphasise that the present work constrains the removal of individual charged grains, but does not yet provide a complete model for the lifetime of a collisional ring with a distribution of grain sizes, due to the limitations we have stated.

Rings around celestial objects are fascinating and complex structures, and have not been observed around moons in the Solar System. We hope that in the future, observations of moons and planets beyond the Solar System will provide opportunities to gain further insights into these systems.

\begin{acknowledgments}
J.M.E. would like to acknowledge funding from a Don Watts Travel Grant, provided by Curtin University.
M.R. gratefully acknowledges support from the Institute for Advanced Study, the Kovner Member Fund, and the Gruber Foundation Fellowship.
We are very grateful to Prof. Oliver Shorttle for helpful discussions, and to an anonymous referee for useful comments that helped us to improve the manuscript.

\end{acknowledgments}

\section*{Data Availability}

The data supporting the findings of this article are available from the authors upon reasonable request. The numerical model was implemented in Python using \texttt{REBOUND} \citep{Rein_2012} and is available on Zenodo under an open-source MIT license: 
\dataset[doi: 10.5281/zenodo.20806732]{https://doi.org/10.5281/zenodo.20806732}.

\appendix

\section{Simulation Results for Titan and Iapetus} \label{A1}

\begin{figure}[ht!]
\centering
\begin{subfigure}{0.47\linewidth}
    \includegraphics[width=\linewidth]{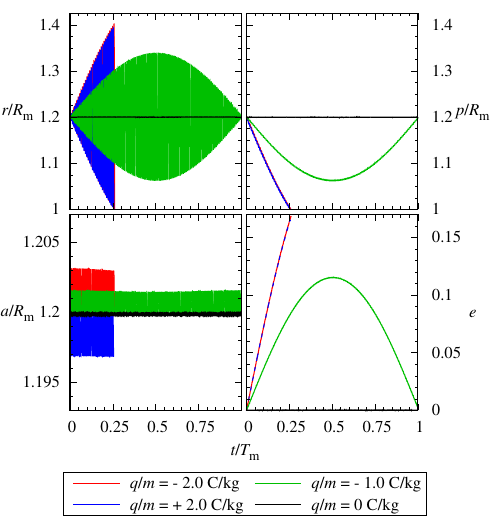}
\end{subfigure}
\hfill
\begin{subfigure}{0.47\linewidth}
    \includegraphics[width=\linewidth]{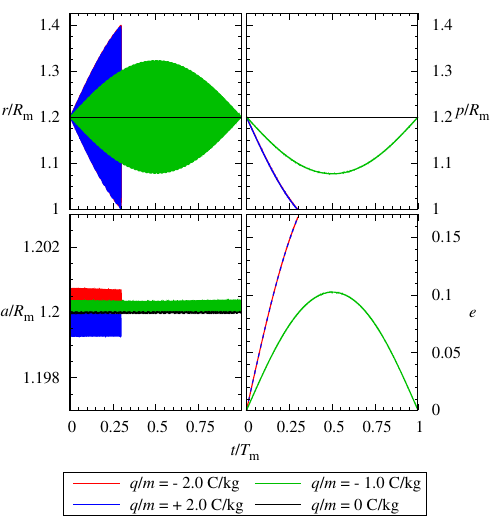}
\end{subfigure}
\caption{Separation distance from moon ($r$), periapsis distance ($p$), semimajor axis ($a$), and eccentricity ($e$) plotted as functions of time for a grain orbiting Titan (left) and Iapetus (right). Our numerical results are presented for grains with charge-to-mass ratios of $-2.0$, $-1.0$, $0$, and $+2.0\,\rm{C/kg}$. \label{fig:Titan_Iapetus}}
\end{figure}

In Fig.~\ref{fig:Titan_Iapetus}, we present simulation results for grains in orbit around Titan and Iapetus. As in Fig.~\ref{fig:Rhea_ts}, the initial orbital radius was chosen to be $1.2\,R_{\rm m}$, and four different charge-to-mass ratios were chosen to show the various dynamical regimes.

\section{Orbital, Physical, and Magnetic Parameters}

The physical and magnetic parameters used in this work are presented in Table~\ref{table:Physical}, and the orbital parameters used are listed in Table~\ref{table:Orbital}. The physical and orbital parameters are the same as those presented in Table~A.1 in \cite{Sucerquia_2024}, which were taken from the NASA Horizons program. The sources of the magnetic dipole moments for the Galilean moons are listed in the table captions.

\begin{table}[ht!]
\begin{center}
\begin{tabularx}{0.54\linewidth}{lccccccc}
\toprule
Body & Mass (kg) & Radius (km) & $T_{\rm{rot}}$ (hours) & $\mu_{\rm p}$ (Tm$^3$) \\ 
 \midrule
 Saturn & $5.68\times10^{26}$ & $6.03\times10^4$ & $10.7$ & $4.6\times10^{18}$ \\
 Rhea & $2.31\times10^{21}$ & $764$ & - & - \\ 
 Titan & $1.35\times10^{23}$ & $2.58\times10^3$ & - & - \\
 Iapetus & $1.81\times10^{21}$ & $734$ & - & - \\ 
 
\bottomrule
\end{tabularx}
\caption{Relevant masses, mean radii, rotation periods, and magnetic dipole moments of the bodies considered in the present study. \label{table:Physical}}
\end{center}
\end{table}

\begin{table}[ht!]
\begin{center}
\begin{tabularx}{0.55\linewidth}{lccccccc}
\toprule
Body & $a$ (\rm{au}) & $e$ & $i$ ($\degree$) & $\Omega$ ($\degree$) & $\omega$ ($\degree$) & $\mathcal{M}$ ($\degree$)  \\ 
 \midrule
 Rhea & $3.53\times10^{-3}$ & $8.70\times10^{-4}$ & $0.489$ & $2.97$ & $3.43$ & $1.42$ \\
 Titan & $8.17\times10^{-3}$ & $0.0287$ & $0.483$ & $2.95$ & $3.07$ & $1.62$ \\
 Iapetus & $0.0238$ & $0.0284$ & $0.298$ & $2.42$ & $4.04$ & $5.28$ \\
\bottomrule
\end{tabularx}
\caption{Orbital elements (semimajor axis ($a$), eccentricity ($e$), inclination ($i$), longitude of ascending node $(\Omega)$, argument of periapsis ($\omega$), and mean anomaly ($\mathcal{M}$)) of Rhea, Titan, and Iapetus used in the present study. \label{table:Orbital}}

\end{center}
\end{table}


\newpage
\bibliography{sample701}{}
\bibliographystyle{aasjournalv7}



\end{document}